\documentstyle[preprint,tighten,aps,floats,epsf]{revtex}
\epsfclipon
\newcommand{\smeq}{\! \! = \!}

\renewcommand{\abstract}{\par
\ifpreprintsty %
\vskip2.5pc
\begin{center}%
{\large \abstractname\par}%
\end{center}%
\vskip-1.0pc
\fi
\bgroup
\ifdim\prevdepth=-1000pt \prevdepth0pt\fi
\hsize\columnwidth
\if@twocolumn\else\leftskip=0.10753\textwidth \rightskip\leftskip\fi
\dimen0=-\prevdepth \advance\dimen0 by17.5pt \nointerlineskip
\small\vrule width 0pt height\dimen0 \relax
}
\renewcommand{\refname}{}
\newcommand{\biblabel}[1]{[#1]} %
\renewcommand{\references}{%
\ifpreprintsty
\vspace*{-0.1 truein}
\hbox to\hsize{\hss\large \refname\hss}%
\else
\vskip24pt
\hrule width\hsize\relax
\vskip 1.6cm
\fi
\list{\biblabel{\arabic{enumiv}}}%
{\labelwidth\WidestRefLabelThusFar  \labelsep4pt %
\leftmargin\labelwidth %
\advance\leftmargin\labelsep %
\ifdim\baselinestretch pt>1 pt %
\parsep  4pt\relax %
\else %
\parsep  0pt\relax %
\fi
\itemsep\parsep %
\usecounter{enumiv}%
\def\theenumiv{\arabic{enumiv}}%
}%
\let\newblock\relax %
\sloppy\clubpenalty4000\widowpenalty4000
\sfcode`\.=1000\relax
\ifpreprintsty\else\small\fi
}

\draft

\newcommand{\bc}{\begin{center}}
\newcommand{\ec}{\end{center}}

\newcommand{\be}{\begin{equation}}
\newcommand{\ee}{\end{equation}}
\newcommand{\ba}{\begin{array}}
\newcommand{\ea}{\end{array}}

\newcommand{\kf}{k_{\scriptscriptstyle F}}
\newcommand{\lambdaf}{\lambda_{\scriptscriptstyle F}}
\newcommand{\Ef}{E_{\scriptscriptstyle F}}

\newcommand{\vf}{v_{\scriptscriptstyle F}}

\newcommand{\br}{{\bf r}}

\newcommand{\G}{{\cal G}}   

\begin{document}


\preprint{25 September 1997, to be published in PHYSICA E}

\title{\bf $ $\\
Semiclassical Approach to Orbital Magnetism of \\
Interacting Diffusive Quantum Systems}

\author{D.~Ullmo,$^{(1,2)}$ K.~Richter,$^{(3)}$ H.U.~Baranger,$^{(1)}$
F.~von~Oppen,$^{(4)}$ and R.A.~Jalabert$^{(5)}$ \\
$ $ }

\address{$^{(1)}$Bell Laboratories--Lucent Technologies, 700 Mountain Ave.,
Murray Hill NJ 07974}

\address{$^{(2)}$Division de Physique Th\'eorique, Institut de Physique
Nucl\'eaire,
91406 Orsay Cedex, France}

\address{$^{(3)}$Max-Planck-Institut~f\"ur~Physik~
komplexer~Systeme,~N\"othnitzer~Str.~38,~01187~Dresden~Germany}

\address{$^{(4)}$Department of Condensed Matter Physics, Weizmann Institute, 
76100 Rehovot, Israel}

\address{$^{(5)}$Universit\'e Louis Pasteur, IPCMS-GEMME,
23 rue du Loess, 67037 Strasbourg Cedex, France}

\maketitle

\mediumtext
{\tighten
\begin{abstract}
We study interaction effects on the orbital magnetism of diffusive
mesoscopic quantum systems. By combining many-body perturbation
theory with semiclassical techniques, we show that the interaction
contribution to the ensemble averaged quantum thermodynamic
potential can be reduced to an essentially classical operator. We
compute the magnetic response of disordered rings and dots for
diffusive classical dynamics. Our semiclassical approach reproduces
the results of previous diagrammatic quantum calculations.
\end{abstract}




}

\vspace*{0.5 truein}


\subsection{Introduction}

The interplay of disorder and interactions in mesoscopic systems has
attracted considerable attention\cite{Efros}.  Interaction effects on
transport through small quantum dots\cite{transport} as well
as on thermodynamic properties like persistent currents and orbital
magnetism are of present interest. In the latter case, the
unexpectedly large measured persistent current of small metal
rings\cite{Levy,Webb1,Webb2} pointed towards the importance of such 
interaction effects and motivated a large number of theoretical
approaches\cite{Efetov}.

For the description of thermodynamic quantities, semiclassical
expansions have proven particularly useful, both within the
independent-particle model\cite{Argaman,vop93,urj95} and for
interaction effects\cite{Montambaux,interball}.  These studies
established a close relation between the classical dynamics and the
quantum-mechanical magnetic response.  In particular, studies of
ballistic systems showed that the quantum thermodynamic properties are
sensitive to whether the classical dynamics is regular or
chaotic\cite{vop93,urj95,interball}.

In this paper we apply these semiclassical techniques
to the orbital magnetism of interacting systems whose non--interacting
classical dynamics is {\em diffusive}. Specifically, we
present semiclassical derivations of the interaction contributions to
the persistent current of metal rings and to the susceptibility of
singly--connected two--dimensional diffusive systems. We
recover results obtained previously by quantum diagrammatic
calculations\cite{aslamazov,AltArZu,doubleA,ambegaokar,Eckern,OhZuSerota}, 
showing that
the semiclassical approach is on the same level of approximation.  By
semiclassically evaluating the relevant diagrams appearing in the
many-body perturbation series for the thermodynamic potential, we
express the latter in terms of an essentially classical operator.
This expression provides a convenient starting point for further
calculations.  Moreover, by making the connection with the classical
dynamics, it provides a physically intuitive picture of the interplay
between disorder and interaction.

\subsection{Diagrammatic perturbation theory}

We are interested in the orbital magnetism of a mesoscopic quantum
system subject to an external magnetic field $B$. While the magnetic
response of a singly--connected system is usually measured in terms of
its susceptibility $\chi$, the magnetic moment of a ring--type
structure threaded by a flux $\phi=B A$ (where $A$ is the enclosed
area) is usually described by the related persistent current $I$.
Both are given in terms of the thermodynamic potential $\Omega$ as
($V$ being the area (volume) of the structure)
\begin{equation}
\label{eq:pcdef}
I \equiv -c\frac{\partial \Omega}{\partial \phi}\quad ; \quad
\chi \equiv - \frac{1}{V} \frac{\partial^2 \Omega}{\partial B^2} \; .
\end{equation}

\begin{figure}
\epsfxsize=12cm
\epsffile[  50 410 550 530 ]{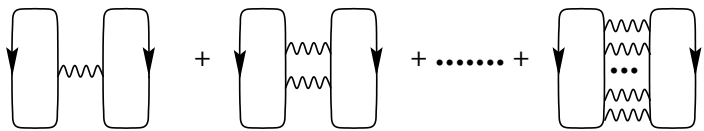}
\caption{
Leading Cooper-channel diagrams for the interaction contribution to the
thermodynamic potential.
}
\label{fig:cooper}
\end{figure}

To calculate the interaction contribution to the magnetic response,
the high-density expansion (RPA) of the thermodynamic potential
\cite{AGD} has to be extended by including interaction corrections
from diagrams with the Cooper-channel.  This was originally performed in
the context of superconducting fluctuations and then applied to
disordered normal 
metals\cite{aslamazov,AltArZu,doubleA,ambegaokar,Eckern,OhZuSerota}.
Such expansions usually yield reliable results even beyond the high
density limit, if the relevant sets of terms are properly resummed.  The
relevant Cooper-like diagrams are shown in Fig.~\ref{fig:cooper}.
The screened Coulomb interaction (wavy lines) can be treated as
local\cite{doubleA,ambegaokar}: $U(\br - \br') = \lambda_0 N(0)^{-1} \delta
(\br - \br')$. Here, $N(0)$ denotes the density of states and the
bookkeeping index $\lambda_0\!=\!1$ identifies the order of
perturbation.  For the local interaction, direct and exchange term are
equivalent up to a factor of $(-2)$ due to the spin sums and the different
number of fermion loops.  The corresponding perturbation expansion for
this interaction contribution $\Omega$ to the thermodynamic potential,
which yields the magnetic response, can be formally expressed as
\cite{AltArZu,doubleA}
\begin{eqnarray}
    \label{omega_diag}
      \Omega & = &  -{1\over \beta} \sum_{n=1}^\infty
       \frac{(-\lambda_0)^n}{n} \,
       \sum_\omega \int d{\bf r}_1 \ldots d {\bf r}_n \,
       \Sigma_{\br_1, \br_2}(\omega) \ldots
       \Sigma_{\br_{n}, \br_1}(\omega)   \\
       & = & {1\over \beta} \sum_\omega {\rm Tr} \left\{
       \ln [1+ \lambda_0\hat \Sigma(\omega) ] \right\} \;.  \nonumber
\end{eqnarray}
Here, $\omega$ denotes the bosonic Matsubara frequencies
$\omega=2 \tilde n \pi / \beta$ with $\beta = 1/kT$.
The particle-particle propagator $\hat\Sigma(\omega)$ is expressed
(in position representation) in terms of products of finite--temperature
Green functions as\cite{AGD}
   \be \label{sigma}
        \Sigma_{\br, \br'}(\omega)={1\over \beta N(0)}
        \sum_{\epsilon}^{\Ef} {\cal G}_{{\bf r}, {\bf r'}}
	(\epsilon) {\cal G}_{{\bf r}, {\bf r'}}(\omega-\epsilon) \; .
   \ee
Here, the sum runs over
the fermionic Matsubara frequencies $\epsilon=(2n+1)\pi / \beta$.
The short-length (high-frequency) behavior is included in the
screened interaction, thus requiring a cutoff of the frequency sums at
the Fermi energy ${\Ef}$\cite{doubleA}.
The straight lines in Fig.~1 represent finite--temperature
Green functions of the non--interacting system. They are of the form
\begin{equation} \label{Gtemp}
\G_{\br, \br'}(\epsilon) = \theta (\epsilon)G^R_{{\bf r},{\bf r'}}
(\Ef \!+\!i\epsilon) + \theta(-\epsilon) G^A_{{\bf r}, {\bf r'}}
(\Ef \!+\!i\epsilon)
\end{equation}
in terms of the retarded and advanced Green functions $G^{R,A}$ which
are related by $G^A_{\br,\br'}(E)=[G^R_{\br',\br}(E^*)]^*$.
For diffusive systems, they include the presence of the
disorder potential.

\subsection{Semiclassical formalism}

Both in ballistic and diffusive samples, the Fermi wavelength
$\lambda_F$ is often the shortest lengthscale. It is in this situation
that we can apply semiclassical techniques to compute $\Sigma_{\br,
\br'}(\omega)$. Here, we will moreover assume that the magnetic field
$B$ is classically weak, i.e., that the cyclotron radius $R_c\gg{\rm
min}\{l,L\}$ (with $l$ the elastic mean free path and $L$ the system
size).

Semiclassically, the retarded Green function is represented as a sum
of contributions $G^{R;j}_{{\bf r},{\bf r'}}$ over all classical paths
$j$ from $\br$ to $\br'$ \cite{LesHqchaos},
\begin{equation}
        \label{GRA} G^R_{\br, \br'}(E) \simeq
        \sum_{j : {\bf r} \to {\bf r'}}
        D_j \, e^{iS_j/\hbar - i\pi\nu_j/2}   \, .
\end{equation}
Here $S_j \smeq \int_{\bf r}^{\bf r'} {\bf p} \cdot d{\bf r}$ is the
classical action of trajectory $j$. The prefactor $D_j$ includes the
classical phase space density [$D_j=(1/\sqrt{2\pi(i\hbar)^3\dot x \dot x'})
\left| {\partial^2 S_j / \partial y \partial y'} \right|^{1/2}$
in two dimensions].  $\nu_j$ is a Maslov index.  The semiclassical
approximation makes the temperature and magnetic-field dependences of
the finite--temperature Green function transparent.  Employing
$(\partial S_j /\partial E) \smeq t_j$ and $(\partial S_j /\partial B)
\smeq (e / c) A_j $, where $t_j$ and $A_j$ are the traversal time and
area, one finds
\begin{equation} \label{Gexpand}
G^{R;j}_{{\bf r},{\bf r'}}(\Ef\!+\!i\epsilon,B)\simeq G^{R;j}_{{\bf r},
{\bf r'}}(\Ef,B\!=\!0)
\times\exp\left[ -\epsilon t_j/\hbar\right] \times \exp
\left[ i 2 \pi B A_j/\phi_0 \right]
\end{equation}
where $\phi_0\!=\!h c / e $ is the flux quantum. Note that temperature
exponentially suppresses the contributions of long paths to each
Green function.

Semiclassically, the particle-particle propagator $\Sigma_{{\bf r},
{\bf r'}}(\omega)$ is then represented as a sum over pairs of paths
between $\br$ and $\br'$.  Off--diagonal pairs (of different paths)
generally contain highly oscillatory contributions which do not
survive an ensemble (disorder) average. (There can be exceptions as
discussed in Ref.\ \cite{interball}.)  On the other hand, the diagonal
pairing of each orbit $j$ with its time reverse persists upon
averaging since their dynamical phases $\exp[i S_j(B\!=\!0)/\hbar]$
cancel while retaining a magnetic-field dependence. A more detailed
semiclassical analysis\cite{interball} shows that the Cooper series in
Fig.~1 contains the magnetic-field sensitive contribution to $\Omega$
which is leading order in $\hbar$.

Using Eqs.~(\ref{Gtemp}), (\ref{GRA}) and (\ref{Gexpand}) in Eq.~(\ref{sigma})
and performing the Matsubara sum yields for the diagonal part
of $\hat{\Sigma}$
\be
\Sigma^{(D)}_{\br, \br'}(\omega)\simeq {\hbar \over
\pi N(0) } \sum_{ j : {\bf r} \to {\bf r'}}^{L_j > \Lambda_0}
|D_j |^2\, {R(2 t_j/ t_T) \over t_j}
\times\exp\left[\frac{ i4\pi B A_j}{\phi_0}\right] \times
\exp\left[-\frac{\omega t_j }{ \hbar}\right]\; .
\label{sigmaD}
\ee
The sum runs over all trajectories longer
than the cutoff $\Lambda_0 = \lambdaf / \pi$ [corresponding to the upper
bound $\Ef$ on the Matsubara sum in Eq.~(\ref{sigma})].
The temperature dependence in Eq.~(\ref{sigmaD}) enters through
the function $R(x) = {x/\sinh(x)}$ introducing the time scale
\be
\label{tT}
t_T = \frac{\hbar\beta}{\pi}
\ee
and the related length scale $L_T = \vf t_T$, with $\vf$ being the Fermi
velocity.
This semiclassical framework allows us to
reduce the original quantum problem to $\Sigma^{(D)}$ which no longer
exhibits variations on the quantum scale $\lambda_F$ but only on
classical scales. We emphasize that the representation (\ref{sigmaD})
of $\Sigma^{(D)}$ is rather general since we have not yet made any
assumption about the classical dynamics of the system. In particular, it
applies to both diffusive and ballistic systems. On the basis of
Eq.~(\ref{sigmaD}), we have recently studied interaction effects in
ballistic quantum dots\cite{interball}.  Specifically, we show
that the interaction induced orbital magnetism scales
differently for systems with regular and chaotic non--interacting
classical counterparts.

Here, we focus on diffusive systems for which it is useful to relate
$\Sigma^{(D)}$ to classical probabilities satisfying the
diffusion equation.  To this end we introduce an additional time
integration in Eq.~(\ref{sigmaD}) and make use of the
relation\cite{Argaman}
\be
\frac{1}{2\pi^2} \, \sum_{ j : {\bf r} \to {\bf r'} }
|D_j |^2 \delta(t-t_j) = \frac{N(0)}{2\pi \hbar} P({\bf r, r'};t)
\ee
between the weights $|D_j|^2$ and the classical probability $P({\bf r,
r'};t)$ to propagate from $\bf r$ to $\bf r'$ in time $t$.

An $n$-th order contribution to $\Omega$ in Eq.~(\ref{omega_diag})
then contains expressions for the joint return probability
$P(\br_1,\ldots,\br_n,\br_1;t_1, \ldots, t_n|A)$ to visit the points
$\br_i$ (with $t_i$ being the time between $\br_i$ and $\br_{i+1}$)
under the condition that the enclosed area is $A$.  In diffusive
systems, the probability is multiplicative, namely $\int d \br_1 \ldots
d \br_n P(\br_1,\ldots,\br_n,\br_1;t_1, \ldots, t_n|A) = \int d \br
P(\br,\br; t_{tot}|A)$ with $t_{tot} = \sum t_i$.  The contribution to
$\Omega$ in Eq.~(\ref{omega_diag}) from the diagonal terms
$\Sigma^{(D)}$ then yields
\newpage
\be
   \label{omega_sc}
 \Omega^{(D)} = \sum_n \Omega^{(D)}_n =
     {1\over \beta} \int d \br \int dt \coth\left(\frac{
     t}{t_T}\right) \, K(t) \, {\cal{A}}(\br,t;B)
\ee
where  the  $\coth(t/t_T)$       arises     from       the  $\omega$--sum  in
Eq.~(\ref{omega_diag}) and
\begin{eqnarray}
\label{KAa}
K(t) & \equiv & \sum_n K_n(t) = - \sum_n \frac{(-\lambda_0)^n}{n} \,
   \left\{ \int  \prod_{i=1}^n \left[ {dt_i R(2 t_i/ t_T) \over t_i}\right]
    \, \delta(t-t_{tot}) \right\} \\
\label{KAb}
{\cal{A}}(\br,t;B) & \equiv & \int d A \cos\left(\frac{4\pi B A}{\phi_0}\right)
      P(\br,\br;t|A)  \, .
\end{eqnarray}
$K(t)$ accounts for temperature effects while $\cal{A}$ contains the
field dependence and the classical return probability.
Eqs.~(\ref{omega_sc})--(\ref{KAb}) are a general and
convenient starting point to compute the orbital response of
disordered systems.

\subsection{Diffusive rings}

We start with the computation of the first-order interaction
contribution, $\Omega^{(D)}_1$, to illustrate the main ideas.
Consider a (thin) disordered ring of width $b$, crossection $\sigma$
and circumference $L$. For $L \gg l,b$ the motion of
particles around the ring effectively follows a law for
one--dimensional diffusion.  Since the area enclosed is given
in terms of the number $m$ of windings around the ring, one has
\be
P(\br,\br;t|A) = \sum_{m=-\infty}^{+\infty}
{1 \over \sigma}
\frac{1}{\sqrt{4\pi Dt}} \exp\left(-\frac{m^2L^2}{4Dt}\right) \,
\delta \left(A - \frac{m L^2}{4\pi}\right) \; ,
\ee
where $D=\vf l/d$ is the diffusion constant (in $d$ dimensions).
Because of the disorder average the classical return probability
does not depend on $\br$.  In first order we have
        \be
        K_1(t)=\lambda_0 R(2t/t_T)/t \, .
        \ee
Combining this with the $\coth$ function in Eq.~(\ref{omega_sc}) we find
	\be \label{omega1_ring}
    \Omega^{(D)}_1 =  \lambda_0 {L \hbar \over \pi} \sum_{m=-\infty}^{+\infty}
   	\cos\left(\frac{4\pi m \phi}{\phi_0} \right) g_m(T) \,
	\ee
with
	\be
	g_m(T) = \int_0^\infty  dt \frac{R^2(t/t_T)}{t^2}
	\frac{\exp\left[{-(mL)^2}/{(4Dt)}\right]}{\sqrt{4\pi Dt}} \, .
	\ee
Taking the derivative with respect to the flux,
we recover the first-order interaction contribution to the persistent
current, first obtained in \cite{ambegaokar} by purely diagrammatic
techniques,
	\be \label{I1_ring}
 	I_1 = \lambda_0{2 L e \over \pi } \sum_{m=-\infty}^{+\infty}
   	m \sin \left( \frac{4\pi m \phi}{\phi_0} \right) g_m(T) \, .
	\ee
Semiclassically, this first order result was already derived by
Montambaux\cite{Montambaux}.

In addition, our semiclassical approach allows us to obtain the 
renormalization
of the coupling constant\cite{aslamazov,AltArZu,doubleA,ambegaokar,Eckern}
due to the higher-order diagrams of the Cooper series.
Including these diagrams amounts to using the full kernel $K(t)$ in
Eq.~(\ref{omega_sc}) instead of $K_1 (t)$.
Introducing the Laplace transform of $K_1(t)$,
\be
\label{hatf} \hat f(p) = 4 \lambda_0 \sum_{n=0}^{n_F} \frac{1}{p t_T
+2(2n+1)}
\ee
($n_F \smeq \beta E_F / 2 \pi \smeq \kf L_T / 4$),
$K(t)$ is given by the inverse Laplace transform
\be \label{renorm}
\label{Kt} K(t) = \frac{1}{2\pi i} \
\int_{-i \infty}^{+i \infty} \!\! dp \, e^{+pt} \, \ln [1 + \hat f (p)]
\simeq \frac{1}{\lambda_0 \ln (\kf L^*)} K_1(t)
\quad ; \quad L^* = \min(\vf t, L_T /4) \; .
\ee
The last equality is valid when $\ln \kf L^* \gg 1$ which is certainly
satisfied when $\ln \kf l \gg 1$.
Therefore, the higher-order terms merely lead to a renormalization of
the coupling constant, thus reducing the predicted magnitude of the 
persistent current.
In the high temperature regime ($L_T \ll L_m$) the coupling
constant is renormalized to ${1}/{ \ln(\kf L_T / 4)}$.
Introducing $L_m \smeq \vf (m L)^2/4 D$, the average length of a trajectory
diffusing $m$ times around the ring, one gets at low temperature
($L_T \gg L_m$) a replacement of $\lambda_0\!\equiv\!1$ by
$1 / \ln(\kf L_m) $.
These two limits agree with results obtained diagrammatically
by Eckern \cite{Eckern}.

We note that the functional form of the temperature dependence
(exponential $T$--damping\cite{ambegaokar}) is in line with
experiments\cite{Levy,Webb1,Webb2} while the amplitude of the persistent
current with renormalized coupling constant is smaller than the experiments
by a factor of $\sim 5$.

\subsection{Diffusive two--dimensional systems}

Contrary to rings, the geometry imposes no shortest length for returning
paths in singly--connected systems.  One therefore
expects a different temperature dependence of the magnetic response.

Consider a two-dimensional singly--connected diffusive quantum dot.
In view of the general renormalization property of diffusive systems
Eq.~(\ref{renorm}), the diagonal part of the thermodynamic  potential
from the entire Cooper series [Eq.~(\ref{omega_sc})] can be written as
\be \label{gen:omegaD}
  \Omega^{(D)} = \frac{1}{\beta} \,
  \int d {\bf r} \int dt \
  \frac{1}{\ln(\kf \, \vf t)} \
  \frac {t_T}{t^2} \  R^2\left(\frac{t}{t_T}\right)
   \ {\cal A}({\bf r},t;B) \, .
\ee
Here we have used $L^* = \vf t$ in (\ref{renorm}) since the $R^2$ factor
ensures that the main contribution to the integral comes from $t < t_T$.
In two dimensions the conditional return probability, entering into
${\cal{A}}$, is conveniently expressed in terms of the Fourier transform
\cite{Argaman}
\be \label{TFreturn}
      P(\br,\br,t|A) = \frac{1}{4 \pi^2} \ \int dk \ |k| \ e^{ikA}
        \frac{\exp(-|k|Dt)}{1 - \exp(-2|k|Dt)}
\ee
from which one obtains
\newpage
\be
\label{A2dim}
	{\cal A}({\bf r},t;B) = \frac{1}{4 \pi D} \frac{R(t/t_B)}{t} \; .
\ee
Here we introduced the magnetic time
\be
\label{tB}
t_B = \frac{\phi_0}{4 \pi B D} = \frac{L_B^2}{4\pi D} \; .
\ee
It is related to the square of the magnetic length $L_B^2$ which
denotes the area enclosing one flux quantum (assuming diffusive dynamics).
Note that the function $R$ in Eq.~(\ref{A2dim})
has a different origin than in Eq.~(\ref{gen:omegaD}).

Using the expression (\ref{A2dim}) in Eq.~(\ref{gen:omegaD}) and taking the
second derivative with respect to the field, we find for the susceptibility,
\be
\label{bulk:chi1}
 \frac{ \chi^{(D)} }{|\chi_{\rm L}|}
     = -\frac{6}{\pi} (\kf l)
	   \int_{\tau_{el}}^\infty
  \frac{d t}{t \ln(\kf\,\vf t)} R^2\left(\frac{t}{t_T}\right)
  R''\left(\frac{t}{t_B}\right)
\ee
where $R''$ is the
second derivative of $R$. The susceptibility is normalized to the
two--dimensional diamagnetic Landau susceptibility
$\chi_{\rm L}= - e^2/(12\pi m c^2)$.

In the above time integral the
elastic scattering time $\tau_{el} = l/\vf$ enters as a lower bound. This
cutoff
must be introduced since for backscattered paths with
times shorter than $\tau_{el}$ the diffusion approximation
(\ref{TFreturn}) no longer holds\cite{cleanbulk}.
On the other hand Eq.~(\ref{bulk:chi1}) holds
true only as long as the upper cutoff time $t^* \! \equiv \! \min(t_T,t_B)$
is smaller than
the Thouless time $t_c \smeq L^2/D$ (with $L$ being the system size).
For times larger than $t_c$ the dynamics begins to behave ergodically, and 
the two-dimensional diffusion approximation is no longer valid.
Assuming $t^* < t_c$, Eq.~(\ref{bulk:chi1})
can be approximately evaluated by replacing $R(t/t_T)$ and $R''(t/t_B)$
by $R(0)=1$ and $R''(0) = -1/3$, respectively, and introducing the upper 
cutoff $t^*$ in the integral.  The remaining
integral yields for $t^* \gg \tau_{el}$
\be
\label{integral}
 \int_{\tau_{el}}^{t^*} \frac{d t}{t \ln(\kf \, \vf t)}
 = \ln \left\{\frac {\ln [\kf \, \vf \min(t_T,t_B) ]} {\ln(\kf l)} \right\}\,.
\ee
The log-log form produced by the $1/t \ln t$ dependence results from the
wide distribution of path-lengths in the system-- there are flux-enclosing
paths with lengths ranging from about $\vf \tau_{el}$ up to $\vf t^*$.
In contrast, in the ring geometry discussed in the previous section the
temperature dependence is exponential because the minimum length of
flux-enclosing trajectories is the circumference.

The averaged susceptibility of a diffusive two--dimensional structure
then reads
\be
\label{chifinal}
 \frac{ \chi^{(D)} }{|\chi_{\rm L}|}
\simeq \frac{2}{\pi} (\kf l)
\ln \left\{  \frac {\ln [\kf \, \vf \min(t_T,t_B) ]} {\ln(\kf l)} \right\} \;.
\ee
One thus finds a log-log temperature dependence
for $t_T < t_B$ and a log-log $B$ dependence  for $t_T > t_B$.
With regard to magnitude, the magnetic  response of diffusive systems is
paramagnetic and enhanced by a factor $\kf l$ compared to the clean Landau
susceptibility $\chi_{\rm L}$.

Eq.~(\ref{chifinal}) agrees with results from Aslamazov and
Larkin\cite{aslamazov}, Altshuler, Aronov and Zyuzin\cite{AltArZu,doubleA},
and Oh, Zyuzin and Serota\cite{OhZuSerota}
obtained with quantum diagrammatic perturbation theory.
The equivalence between the semiclassical and quantum approaches to
diffusive systems may be traced back to the fact that the ``quantum''
diagrammatic perturbation theory relies on the use of the small
parameter $1/\kf l$ which can be viewed as a semiclassical approximation.

\subsection{Conclusion}

To conclude, we developed a semiclassical approach to evaluate
the interaction contribution of the grand potential in a high density
perturbative expansion. We showed that the averaged quantum magnetic
response can be expressed in terms of an operator containing the
classical probability for particles to return. As an application we
computed the orbital magnetic response of diffusive rings and
two--dimensional quantum dots arising from the combined effects of
disorder and interaction.

RAJ and KR acknowledge support from the
French-German program PROCOPE. The Division de Physique Th\'eorique is
``Unit\'e de recherche des Universit\'es Paris~11 et Paris~6 associ\'ee au
C.N.R.S.''.


\begin{references}
\bibitem{cleanbulk} Shorter paths with $t < \tau_{el}$ arise
                    from higher order interaction events and contribute
                    to the clean bulk magnetic response which however is
                 negligible compared to the disorder induced interaction
                  contribution\cite{interball}.
\end{references}
\end{document}